\documentclass{PoS}
\usepackage{graphicx,pstricks,epsfig,rotating,amsmath}

\title{Interplay between Symmetry Energy and Excluded Volume Corrections under the 
Direct Urca Cooling Constraint in Neutron Stars}

\ShortTitle{Excluded volume, $E_{s}(n)$ and DU cooling in NS}

\author{\speaker{David E. Alvarez-Castillo}\\
        Bogoliubov  Laboratory of Theoretical Physics,
JINR Dubna, 141980 Dubna, Russia\\
Instituto de F\'{\i}sica,
Universidad Aut\'onoma de San Luis Potos\'{\i},
S.L.P. 78290, M\'exico
\\
        E-mail: \email{alvarez@theor.jinr.ru}
}

\author{David Blaschke\\
        Institute for Theoretical Physics,
University of Wroc{\l}aw,
50-204 Wroc{\l}aw, Poland\\
Bogoliubov  Laboratory of Theoretical Physics,
JINR Dubna, 141980 Dubna, Russia
\\
National Research Nuclear University (MEPhI),
115409 Moscow, Russia
\\
        E-mail: \email{blaschke@ift.uni.wroc.pl}}

\abstract{Both the symmetry energy part and excluded volume corrections to the equation of state play an important role for the neutron star interior structure and composition, namely for the profile of the baryon density and the proton fraction.
While the symmetry energy uniquely determines the proton fraction, excluded volume effects control the maximum density values inside neutron stars. 
Observations of cooling neutron stars indicate that the fast direct Urca cooling is not operative for the typical, low mass stars, pointing at proton fractions that lie
below the threshold for the onset of direct Urca cooling process. 
This in turn, restricts the density range admissible in neutron star interiors and may require an excluded volume correction. 
In this contribution we discuss the interplay between fast cooling, symmetry energy and excluded volume corrections to the equation of state that would be required to fulfil the direct Urca cooling constraint.}

\FullConference{The Modern Physics of Compact Stars 2015\\
		30 September 2015 - 3 October 2015\\
		Yerevan, Armenia}

\newcommand{\bea}{\begin{eqnarray}}
\newcommand{\eea}{\end{eqnarray}}
\newcommand{\apgt} {\ {\raise-.5ex\hbox{$\buildrel>\over\sim$}}\ }

\begin{document}


\section{Introduction}

Astronomical observations of neutron stars (NS) provide a way to test our knowledge of nuclear matter 
properties at high densities such as the equation of state (EoS). 
They allow to derive constraints on their theoretical description by microscopic approaches. 
Basic quantities like mass, radius, moment of inertia, cooling rates are of special interest for this purpose. Of great importance is the density dependent nuclear symmetry energy $E_s(n)$, a quantity that 
determines the properties of isospin asymmetric atomic nuclei and, of course, the composition of neutron star interiors. 
It is best determined from isobaric analog states (IAS) of nuclei~\cite{Danielewicz:2013upa} at the reference 
density $n=n^*\sim 2/3~n_0$ where the saturation density $n_0=0.15$ fm$^{-3}$ is the value one would extrapolate for atomic nuclei of infinite size.
Other laboratory experiments for determining the nuclear symmetry energy include measurements of the nuclear polarizability~\cite{Roca-Maza:2015eza}, parity violation in electron scattering on nuclei \cite{Ban:2010wx}, but also measurements of nuclear masses~\cite{Kortelainen:2011ft}, giant dipole resonances~\cite{Trippa:2008gr} and others (for an exhaustive review, see~\cite{Lattimer:2012xj,Li:2014oda}). 

For pure hadronic neutron stars (composed of neutrons, protons, electrons and muons only), it is the symmetry energy functional that determines the proton content in their interiors. 
Moreover, the proton fraction $x$ can trigger a rapid cooling by the direct Urca (DU) process 
$n\to p + e^- + \bar{\nu}_e$ 
if it overcomes the threshold $x_{\rm DU}$ for the onset of this process~\cite{Lattimer:1991ib}. 
Observational evidence~\cite{Popov:2004ey} shows that typical NS (with masses in the range of 
$1.4 \dots 1.6~$M$_{\odot}$) do not cool by this extremely fast process, since otherwise one could not explain the observation of a broad spectrum of surface temperatures for neutron stars with an age exceeding $\sim 10^3$ years. 
This demonstrates the importance of the DU cooling process as a constraint to the EoS of NS. For a detailed discussion of the DU constraint, see \cite{Blaschke:2006gd,Popov:2005xa,Klahn:2006ir}. 
In this contribution, we consider a one-parameter set of symmetry energy functionals and explore how the restrictions on the variation of that parameter stemming from the DU constraint could be relaxed when allowing for \textit{excluded volume corrections to the nuclear EoS} at supersaturation densities.

\section{$E_s(n)$ functional models}
\label{sec:esym}

In order to describe infinite nuclear matter in neutron star interiors we consider the energy per particle as a functional of the baryon number density $n=n_p+n_n$ and isospin asymmetry $\alpha = (n_n-n_p)/n$
\begin{equation}
E_{\rm nuc}(n,\alpha)=E_{0}(n) + E_s(n)\, \alpha^2 + {\cal O}(\alpha^4)~,
\label{Enuc}
\end{equation}
which in an isospin symmetric system reduces to $E_{0}(n)=E(n,\alpha=0)$. 
Introducing the proton fraction $x=n_p/n$ the asymmetry can be written as $\alpha = (1-2x)$. 
In the following we consider stellar matter that consists of nucleons and leptons and we shall keep 
in (\ref{Enuc}) only the quadratic asymmetry dependence defined by $E_s(n)$ while neglecting higher order corrections, which usually is an excellent approximation.

\subsection{Energy of symmetric matter}

The model we choose in this work for the description of the energy $E_0(n)$ of uniform, infinite, symmetric nuclear matter is based on a generalized relativistic density functional approach using the "DD2" parametrization given in Ref.~\cite{Typel:2009sy}. 
It fulfills all basic constraints from nuclear structure. 
It is also in accordance with the rather stringent constraint that the predicted maximum NS mass should be not less than that of the most massive pulsars for which masses $M \sim 2~$M$_{\odot}$ have been observed~\cite{Demorest:2010bx,Antoniadis:2013pzd}.

\subsection{Symmetry energy}

In this study we discuss two classes of symmetry energy functionals that both are in agreement with constraints from nuclear structure and that have already been successfully applied to the description of neutron stars, see also Ref.~\cite{Blaschke:2016lyx}. 

The first one is obtained from extending the DD2 model by varying the density dependence of the $\rho$-meson coupling to nucleons and then extracting the resulting symmetry energy functional $E_s(n)$. 
The variations defined in \cite{Typel:2014tqa} were applied there for studies of the neutron skin thickness of heavy nuclei and of core-collapse supernova evolution \cite{Fischer:2013eka}.
They predict a sufficiently gentle rise of $E_s(n)$ at supersaturation densities which is not in conflict with the DU constraint \cite{Blaschke:2016lyx}.  
Therefore we do not consider the DD2 based class of symmetry energies further in this contribution.
\begin{figure}[!thbp]
\includegraphics[width=0.65\textwidth, angle=0]{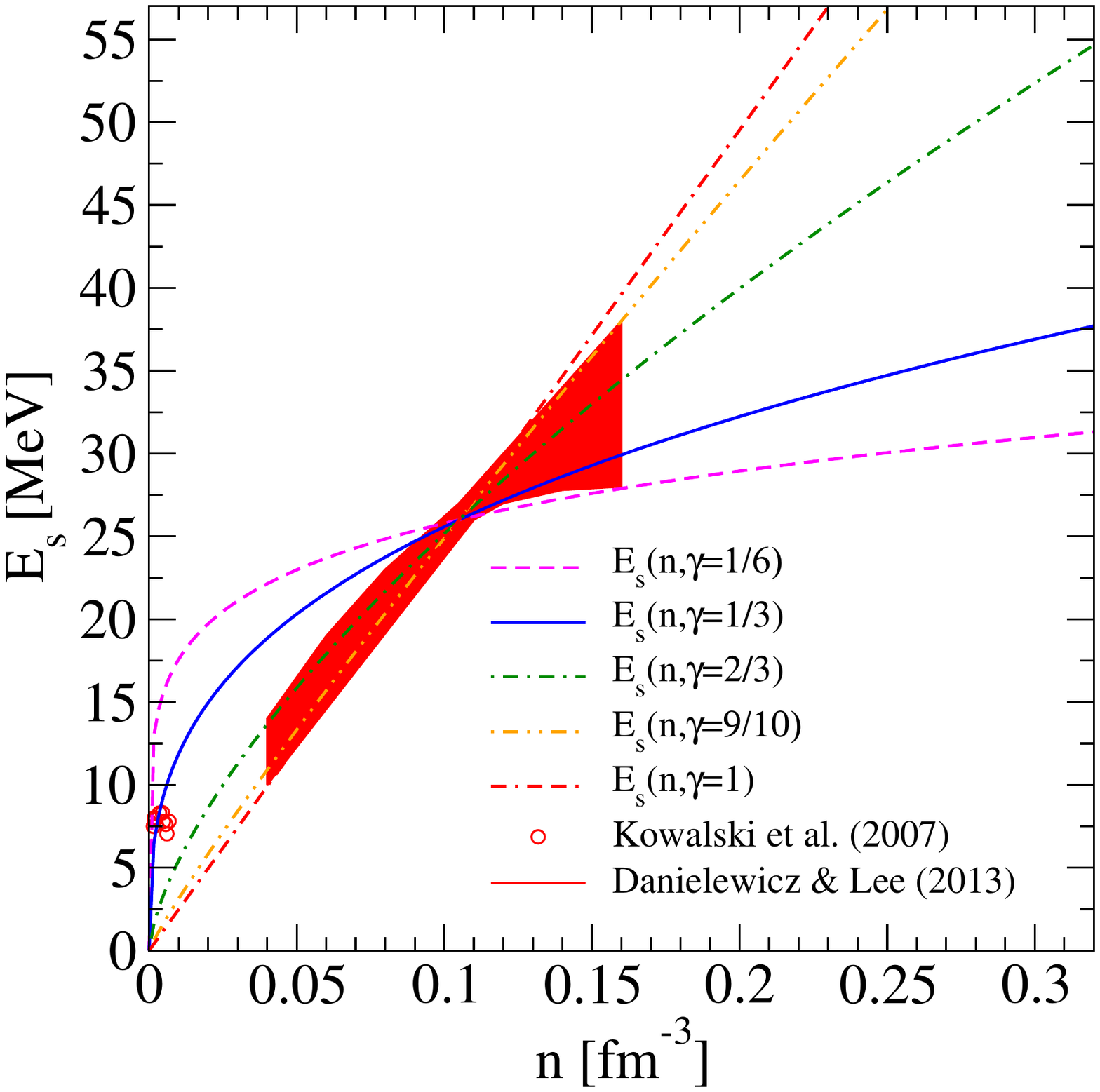}
\hspace{-2.5cm}
\includegraphics[width=0.65\textwidth, angle=0]{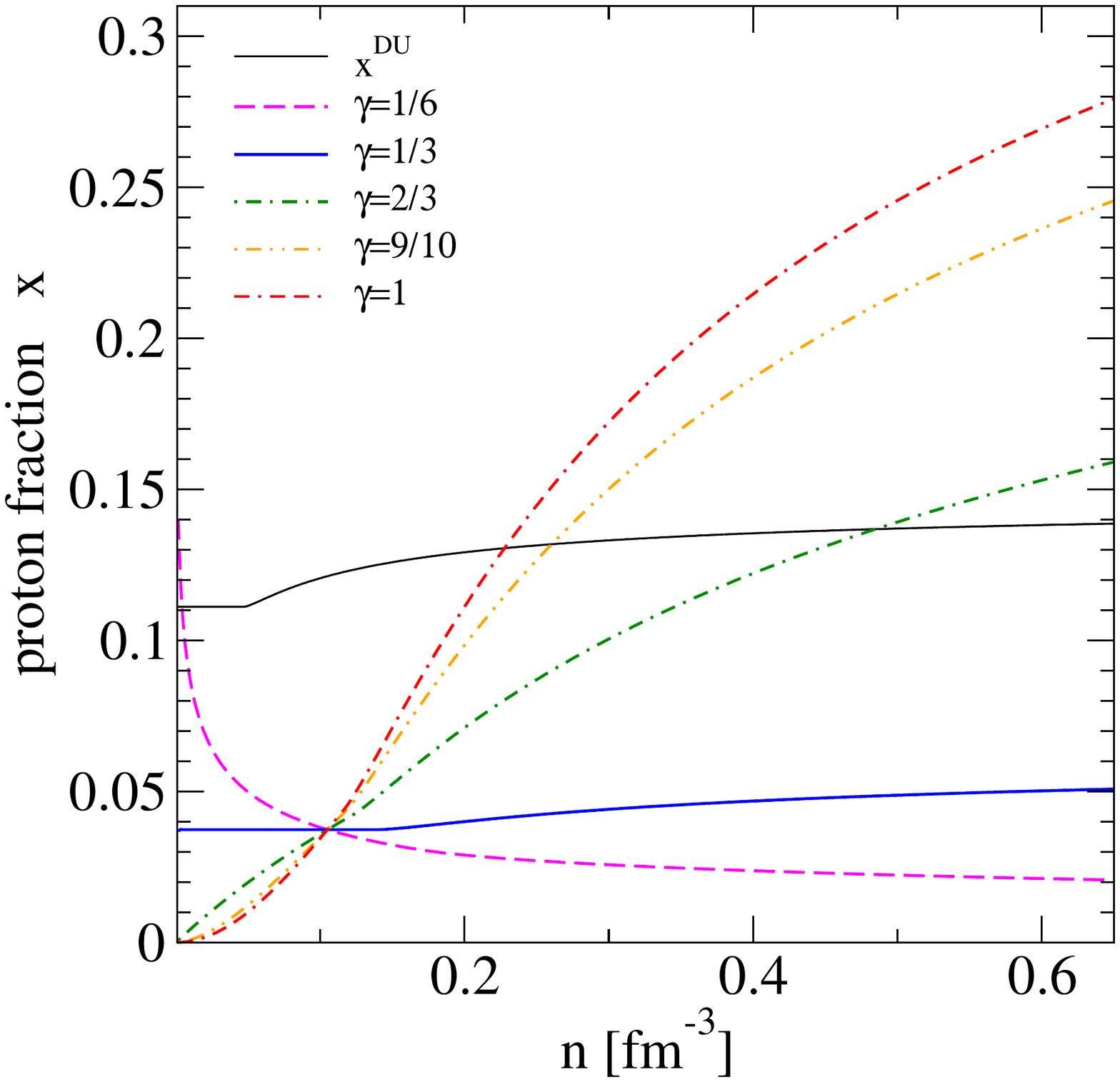}
\caption{{\it Left panel:} Symmetry energy as a function of baryon density for the MDI-type ansatz (\protect\ref{MDI}) compared to the IAS constraint \cite{Danielewicz:2013upa}. Red circles 
located in the low density region correspond to an experimental determination of the symmetry energy 
in collisions of asymmetric nuclei~\cite{Kowalski:2006ju,Hagel:2014wja}.
{\it Right panel:} Proton fraction as a function of baryon density for the MDI-type ansatz (\protect\ref{MDI})
compared to the proton fraction threshold of the direct Urca process.
\label{UEsMuons}}
\end{figure}

To study the effects of the symmetry energy at high densities we base our description on the 
behaviour following from the MDI-type \cite{Tsang:2003td,Tsang:2008fd} power-law ansatz 
\bea
\label{MDI}
E_s(n) = E_s^* (n/n^*)^\gamma~.
\eea
The three parameters are fixed from constraints obtained recently by Danielewicz and Lee \cite{Danielewicz:2013upa} by analysing IAS of nuclei. 
The reference density is found to be $n^*=0.105$ fm$^{-3}$ where the symmetry
energy is $E_s^*=E_s(n^*)=25.7$ MeV.
According to the IAS constraint  \cite{Danielewicz:2013upa} there is a rather narrow error band in the density range $0.04 < n/{\rm fm}^{-3}<0.16$ shown by the red, shaded area in the left panel of 
Fig.~\ref{UEsMuons}.
In the same panel we show the symmetry energy (\ref{MDI}) 
where the exponent $\gamma$ is chosen in the range $1/6 < \gamma < 1$ .
We would like to note that recently the fusion probabilties measured in experiments for the synthesis of superheavy elements were used for the first time to constrain the density dependence of the 
MDI-type symmetry energy employed in their theoretical description to the narrower region 
$1/2 < \gamma < 1$ \cite{Veselsky:2016bjt}.

\begin{figure}[!htb]
\begin{center}
\includegraphics[scale=0.25]{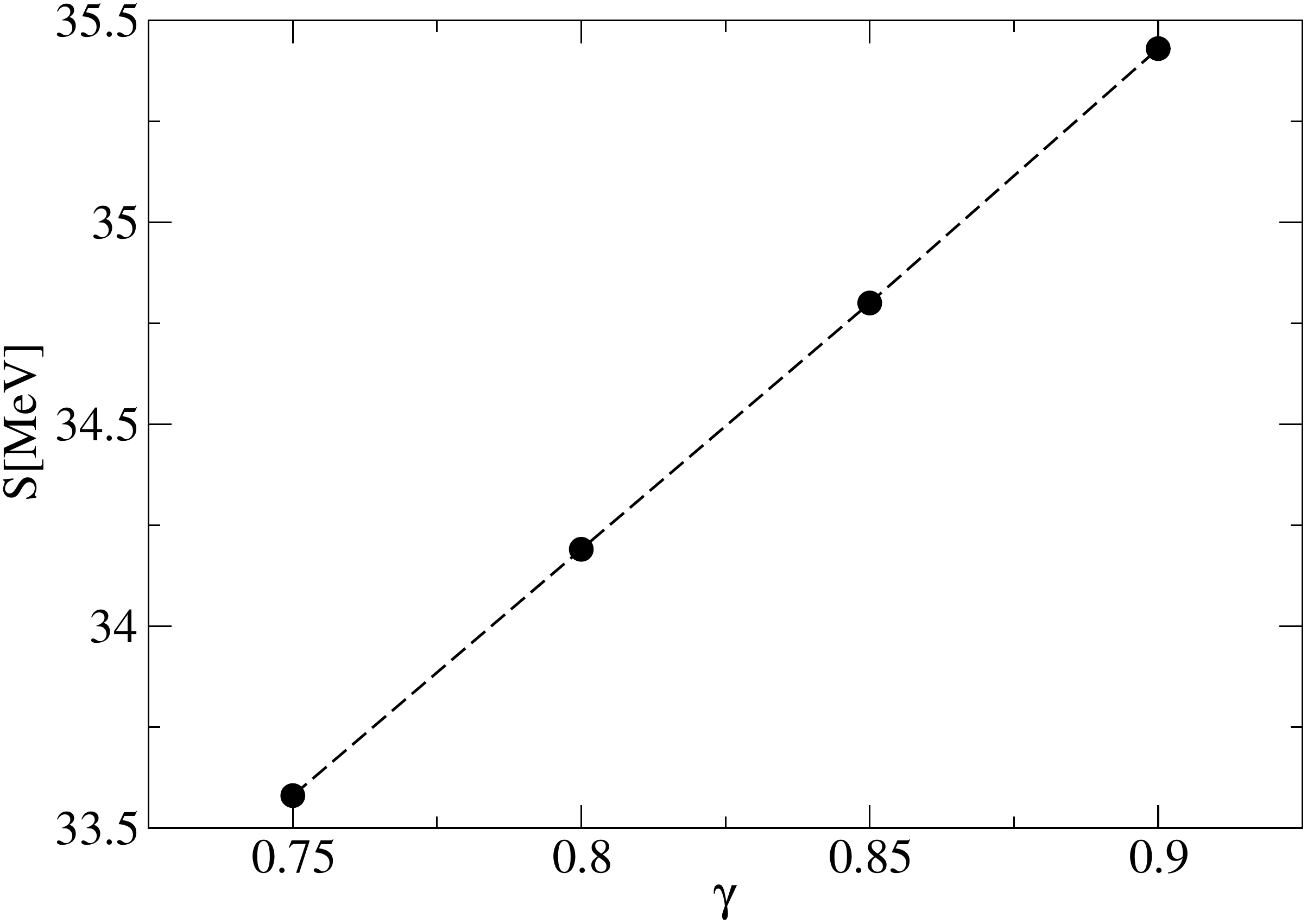}
\hfill
\includegraphics[scale=0.25]{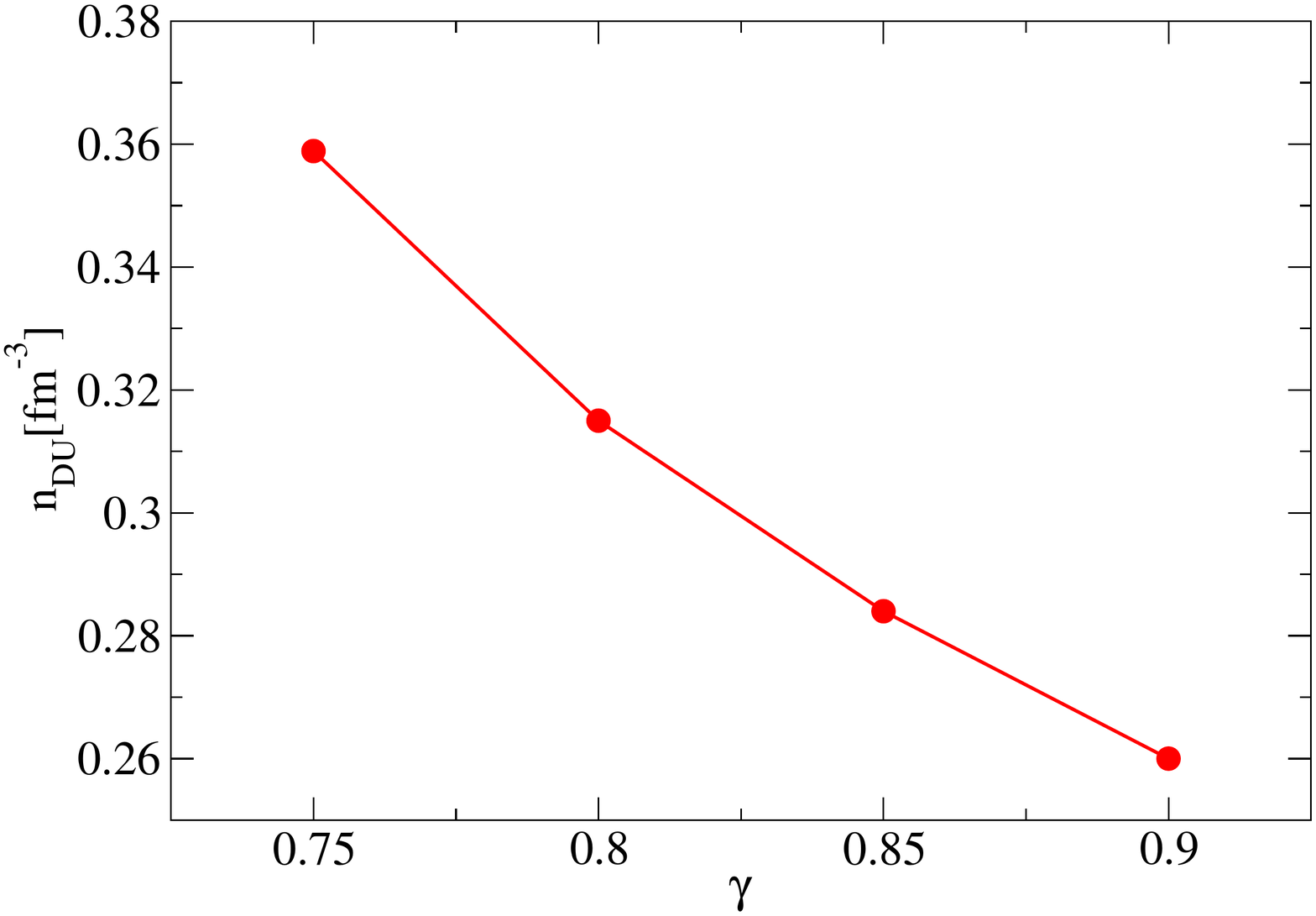}
\end{center}
\caption{MDI-type symmetry energy properties. 
{\it Left panel:} Dependence of the symmetry energy $S=E_s(n_0)$ on the parameter $\gamma$. 
{\it Right panel:} Threshold density $n_{\rm DU}$ for the onset of the DU cooling process as a function 
of the parameter $\gamma$. The $n_{\rm DU}$ value is solely determined by $E_s(n)$ in NS matter.
\label{nc_gamma_S}}
\end{figure}

The symmetry energy parameter is defined as $S=E_s(n)|_{n=n_0}$, and due to the proximity of the saturation density $n_0$ and the reference density $n^*$
an approximately linear dependence for $S(\gamma)$ is obtained, see the left panel of 
Fig.~\ref{nc_gamma_S}.  

\subsection{Direct Urca process constraint}

The MDI-type symmetry energy functional introduced in the previous section serves to solve for the proton and muon fractions under neutron star constraints (for details see, e.g., \cite{Blaschke:2016} or the textbook ~\cite{Haensel:2007yy}). 
As it has been already mentioned, the proton fraction plays an important role in the neutron star phenomenology, in particular since it dictates whether the DU process $n\to p + e + \bar{\nu}_e$ 
can occur or not \cite{Lattimer:1991ib}.
Once the NS central density $n_c$ exceeds the critical value $n_{\rm DU}$ for which holds that
$x(n_{\rm DU})=x_{\rm DU}(n_{\rm DU})$, this fastest neutrino cooling process starts operating and
causes a dramatic drop of the NS core temperature. 
Typical timescales for the cooling wave from the core to reach the surface are about 100 years.
Then, for the case of DU cooling, the photon luminosity of the star drops down rapidly, turning the star practically invisible. 
We do not expect this process to be operative in typical, not too massive neutron stars since we observe cooling neutron stars much older than 1000 years with surface temperatures that are not compatible with this fast DU cooling (see~\cite{Blaschke:2006gd,Klahn:2006ir} for a detailed discussion on the DU process constraint).
To understand the DU threshold condition we look at the triangle inequality for the Fermi momenta of neutron, proton and electron involved in the process 
(neutrino momenta can be safely neglected since they are orders of magnitude smaller) 
leading to the condition
\bea
n_n^{1/3} < n_p^{1/3} + n_e^{1/3}~,
\eea
which (using charge neutrality: $x=x_p=x_e+x_\mu$, $x_i=n_i/n$) is formulated in terms of proton and muon fractions as $(1-x)^{1/3}<x^{1/3}+ (x-x_\mu)^{1/3}$ which is equivalent to
\bea
\label{DU}
x>1/\{1+(1+(n_e/(n_e+n_\mu))^{1/3}]^3\}~.
\eea
For low densities where electrons are the only lepton species one easily recovers the classical result for the DU threshold without muons $x_{\rm DU}=1/9=11.1\%$ \cite{Lattimer:1991ib}.
The density dependence of the DU threshold is obtained by inserting the density dependent muon fraction $x_\mu(n)$ into Eq.~(\ref{DU}); the result is shown in Fig.~\ref{UEsMuons} (right panel). 
The stiffness of the symmetry energy for the MDI-type models is strongly dependent on the $\gamma$ parameter. 
We can see from the right panel of Fig.~\ref{nc_gamma_S} that for all $\gamma \ge 3/4$ the DU constraint is violated  already for densities below $\sim 2.4~n_0=0.36$~fm$^{-3}$. 
The question arises: What values of $n_c$ can one meet in NS of a certain mass not exceeding a typical value? 
The answer to this question shall depend on the neutron star EoS, in particular on its stiffness. 
In order to systematically study this dependence, we will consider a stiffening of the high-density EoS 
by an excluded volume for baryons.    

\subsection{Excluded volume correction to the EoS}

Our choice of EoS for the neutron star description is the density dependent relativistic meanfield EoS 
"DD2" \cite{Typel:2009sy,Typel:1999yq}. 
We implement the excluded volume correction as it is a means to account for an effect of the quark substructure of nucleons which leads to the repulsive quark Pauli blocking interaction in hadronic matter
\cite{Ropke:1986qs,Blaschke:1988zt}. 
This correction as described in Ref.~\cite{Typel:2016srf} results in a stiffening the EoS at supersaturation densities while remaining in accordance with experimental constraints below and around saturation. 
The details of the formulation of the excluded volume modification of this class of EoS models can be found 
in Ref.~\cite{Typel:2016srf}. 
Here we use the available volume fraction $\Phi(n)=1-\mathrm{v_{ex}} n$ with the excluded volume parameter $\mathrm{v_{ex}}=1/n_x$, where $n_x$ is the closest packing density. 
This approach has already been implemented in the description of the hadronic part of hybrid star EoS, 
as presented in~\cite{Benic:2014jia,Alvarez-Castillo:2015rwi,Alvarez-Castillo:2016oln}. 
The possible consequences of implementing a stiffening of the EoS at supersaturation densities
in core-collapse supernova simulations has recently been studied in Ref.~\cite{Fischer:2016jkc}.

It is important to mention that for finite $n_x$ the hadronic EoS becomes acausal at densities $n \sim n_x$. 
This problem, however, is cured in a physical way by allowing for a phase transition to quark matter in the NS interior which is another inevitable consequence of the quark substructure of baryons. 
Furthermore, when the excluded volume correction is introduced in the description of hybrid 
star configurations, this approach serves to solve several problems like masquerades, reconfinement of quark matter at high densities and, the hyperon puzzle, as described in~\cite{Blaschke:2015uva}.

\subsection{The neutron star mass-radius and mass-central density relations}

Above we have defined a family of EoS (\ref{Enuc}) for asymmetric nuclear matter with two free 
parameters: $\gamma$ for the stiffness of the symmetry energy $E_s(n)$ and ${\rm v_{ex}}$ for the stiffness of the EoS of symmetric matter $E_0(n)$\footnote{Our assumption that the excluded volume is isospin independent could be relaxed in a refined calculation which would the modify the results in Figs. 3-5 quantitatively.}.
When augmented with the EoS for leptons $E_{\rm lep}(n)$ and under fulfilment of the NS constraints 
(for which electric charge neutrality and $\beta$-equilibrium conditions determine the density dependent particle fractions $x(n)$, $x_e(n)$ and $x_\mu(n)$) 
\bea
E(n) = \varepsilon(n)/n = E_{\rm nuc}(n) + E_{\rm lep}(n)~,~~ P(n)=n^2 (dE(n)/dn)~.
\eea
With these inputs the NS structure can be obtained by solving 
the  Tolman-Oppenheimer-Volkoff (TOV) equations~\cite{Tolman:1939jz,Oppenheimer:1939ne}  
for the hydrodynamic stability of static, spherically symmetric bodies in the framework of general realitivity
\bea
\frac{dP( r)}{dr}&=& - \frac{G M( r)\varepsilon( r)}{r^2}\frac{\left[1+{P( r)}/{\varepsilon( r)}\right]
\left[1+ {4\pi r^3 P( r)}/{M( r)}\right]}{\left[1-{2GM( r)}/{r}\right]},\\
\frac{dM( r)}{dr}&=& 4\pi r^2 \varepsilon( r) .
\eea
For each initial value of central (energy) density $n_c=n(r=0)$ upon Runge-Kutta integration of these differential equations one gets a pair of values for the mass $M$ and the radius $R$ of the corresponding NS configuration. 
Varying $n_c$ in the appropriate range for a given parametrization of the EoS one obtains a sequence of NS configurations defining uniquely a function $M( R)$, or equivalently $M(n_c)$. 
This equivalence of $P(\varepsilon)$ and $M( R)$ is the theoretical basis for determining the EoS of NS from sufficiently accurate simultaneous measurements of their masses and radii.  
For a recent discussion and present status of this issue, see \cite{Alvarez-Castillo:2016oln} and references therein.

\begin{figure}[!thb]
\includegraphics[width=0.65\textwidth, angle=0]{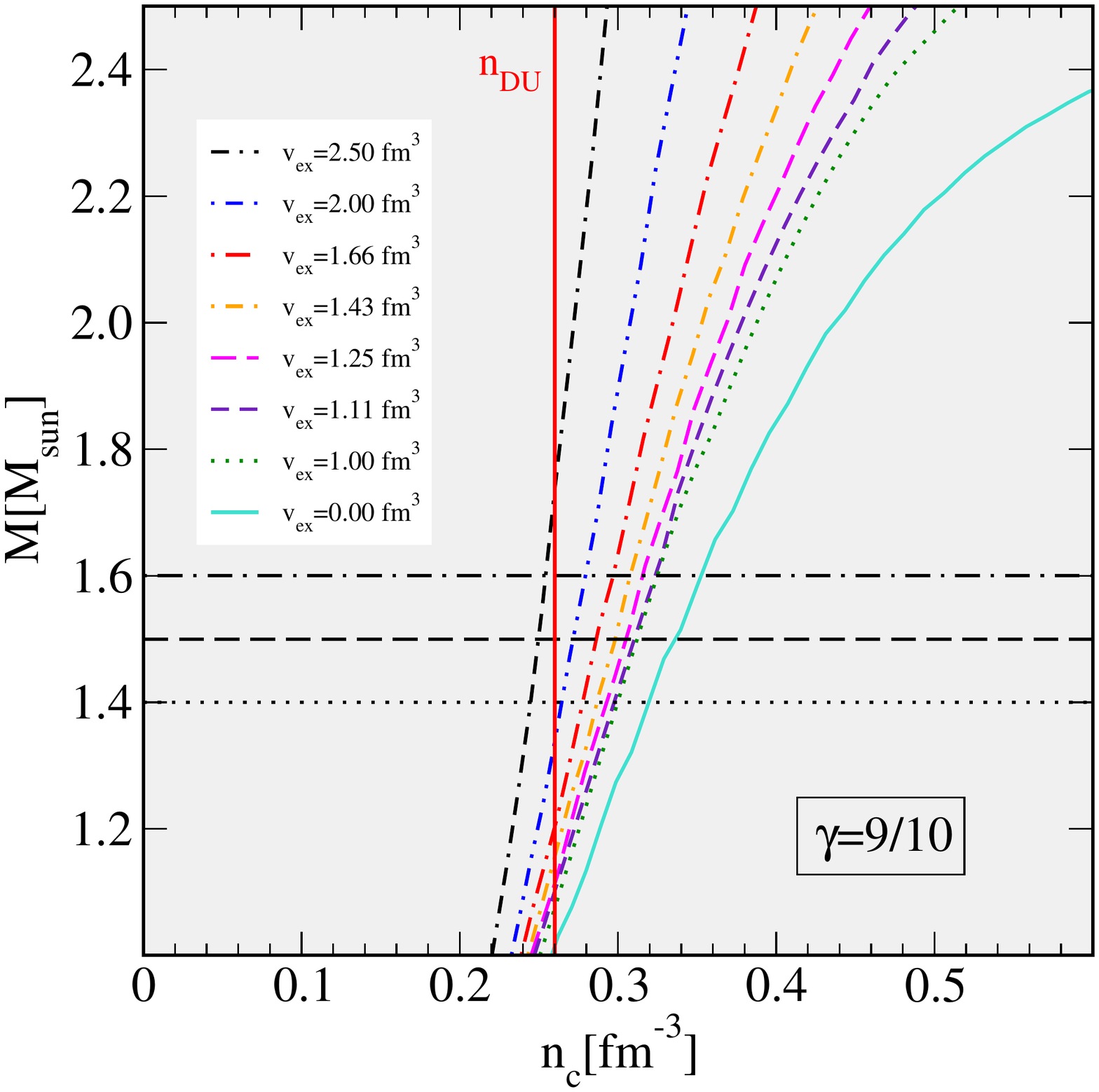}
\hspace{-2.2cm}
\includegraphics[width=0.65\textwidth, angle=0]{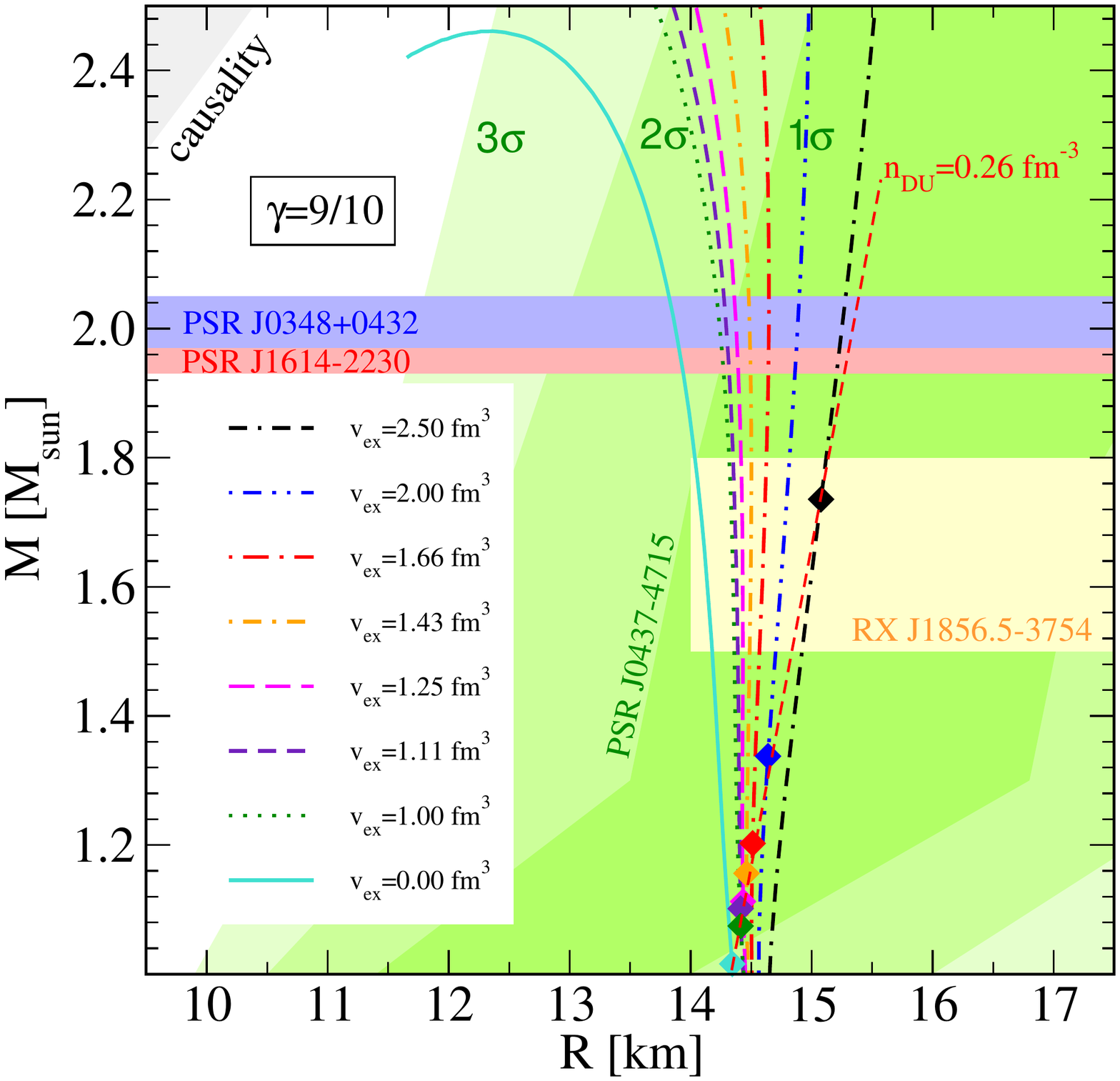}
\caption{Solutions of the TOV equations for a MDI-type $E_s(n)$ model with $\gamma=9/10$ for different values of the nucleon excluded volume ${\rm v_{ex}}$. 
{\it Left panel:} NS mass as a function central density $n_c$. The vertical line marked as $n_{\rm DU}$ corresponds to the threshold density for the onset of DU cooling, which is reached in the centre of a $M=1.6$ M$_\odot$ star for all $\mathrm{v_{ex}}$ values but the highest. 
{\it Right panel:} NS mass-radius relation. The dashed line joining all the diamond symbols 
for each mass-radius curve corresponds to $n_{\rm DU}$. 
Neutron stars with $M<1.6~$M$_{\odot}$ do not suffer from DU cooling only when $\mathrm{v_{ex}} > 2.4$ fm$^{3}$ .
\label{M-R-nc}}
\end{figure}
In Fig.~\ref{M-R-nc} we show the results of the TOV solution using the rather stiff high-density symmetry energy parametrization with $\gamma=9/10$ for different values of the nucleon excluded volume 
$\mathrm{v_{ex}}$. 
On the left panel we demonstrate how stiffening the EoS by increasing $\rm{v_{ex}}$ lowers the central density for a star of given mass and increases the minimal star mass for which the DU process will become operative. These masses and corresponding radii are shown by the diamond symbols in the right panel of that figure.

\begin{figure}[!thpb]
\begin{center}$
\begin{array}{cc}
\includegraphics[width=0.5\textwidth, angle=0]{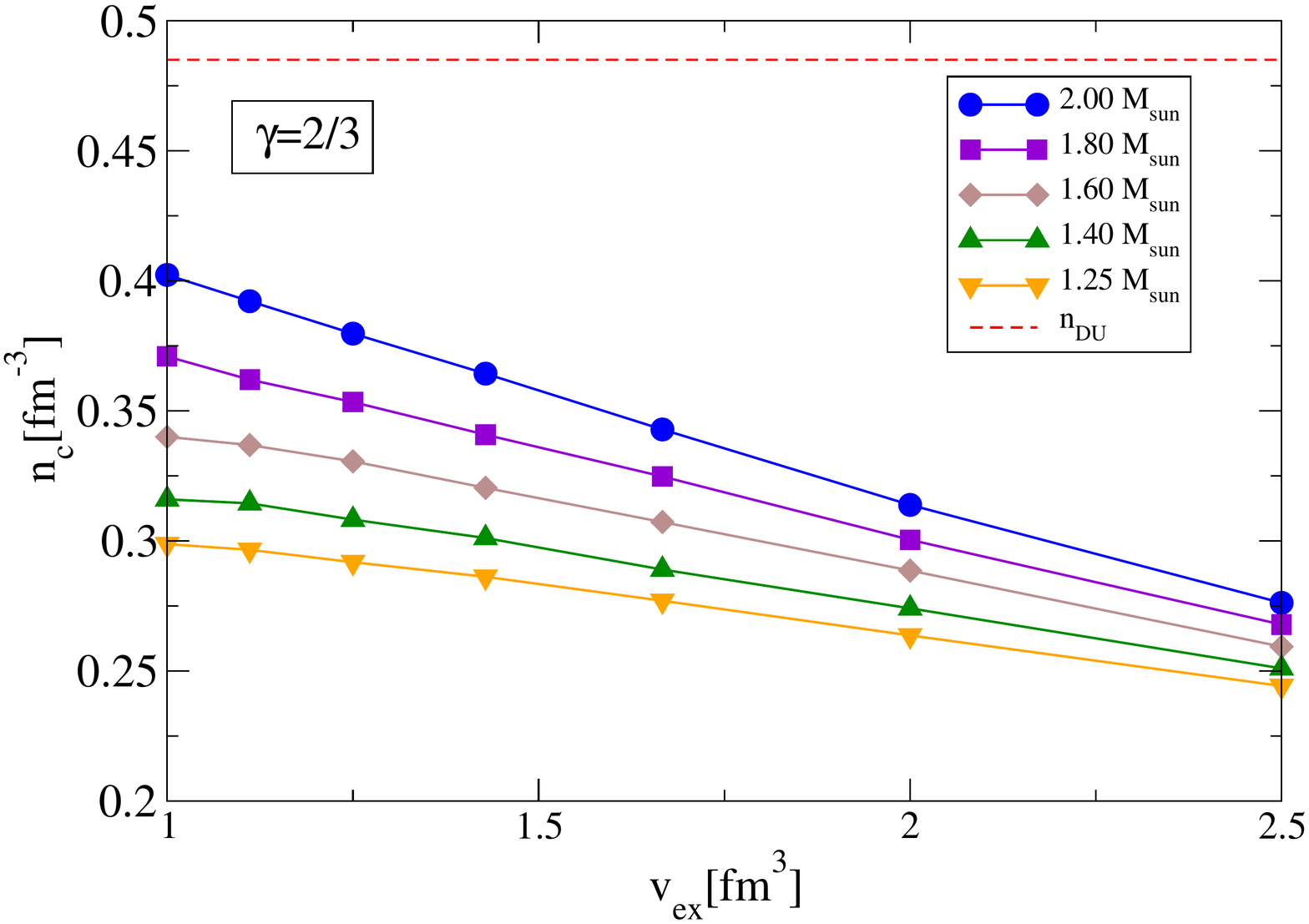}&
\includegraphics[width=0.5\textwidth, angle=0]{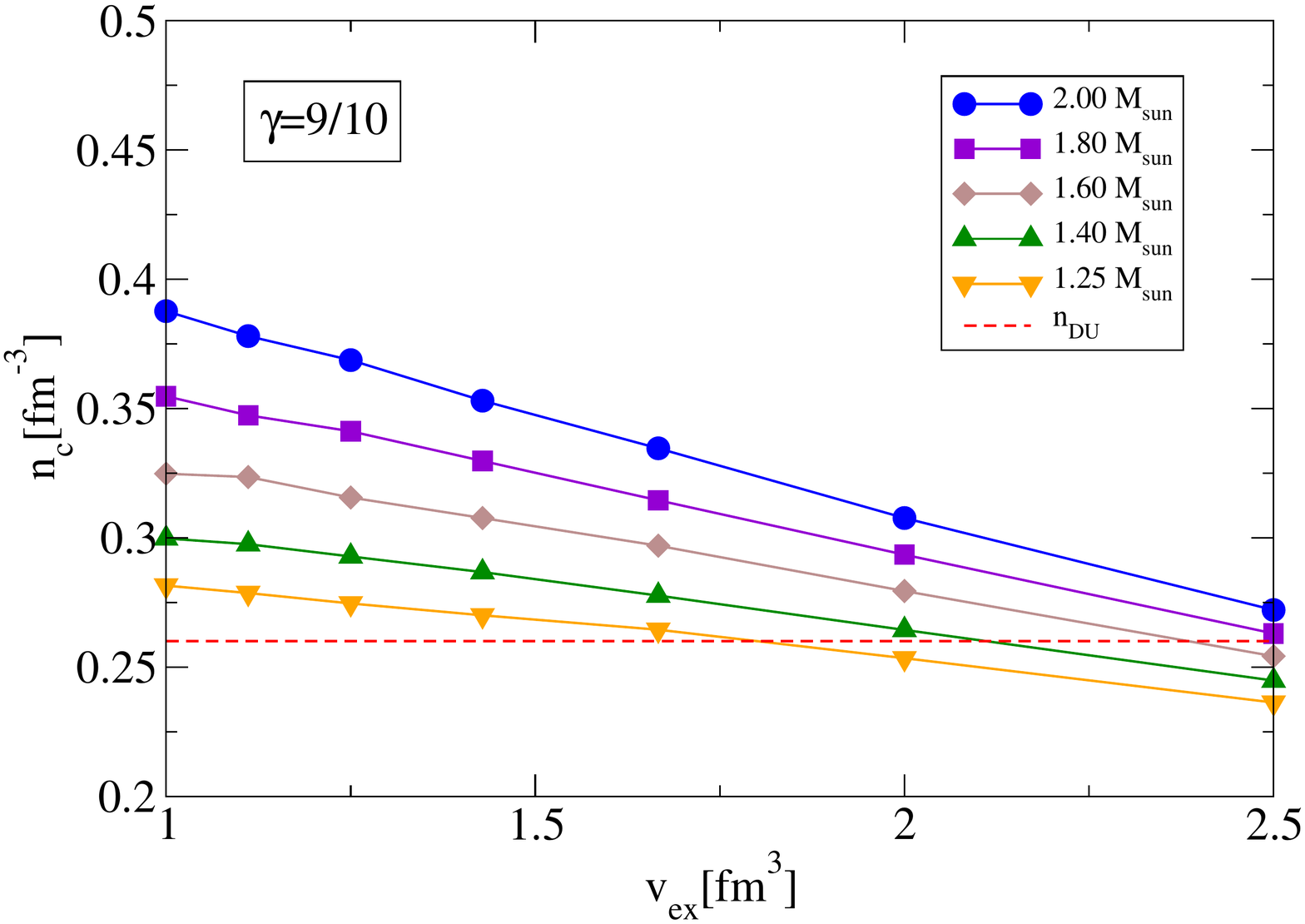}\\
\end{array}$
\end{center}
\caption{Central density versus excluded volume for stars of different mass, compared to the 
threshold density (red solid line) for the direct Urca process.
\label{vx-nc}}
\end{figure}

In the present work we are interested in the consequences that the additional DU constraint 
could have on the stiffness of $E_{s}(n)$ and $E_0(n)$ in the EoS (\ref{Enuc}).
In Fig.~\ref{vx-nc} we show how the central densities of neutron stars depend on the stiffness of the symmetric part of the EoS  for five neutron star masses: 
$1.25,~1.40,~1.60,~1.8$, and $2.0~$M$_\odot$ 
as a function of the excluded volume parameter $\mathrm{v_{ex}}$ which controls the high-density 
stiffening of the $E_0(n)$ part for the employed DD2 EoS.
The two panels of Fig.~\ref{vx-nc} differ by the  MDI-type symmetry energy used:
$\gamma=2/3$ in the left panel and $\gamma=9/10$ in the right panel.
It is clear that for low $\gamma$ values like  $\gamma=2/3$  (which exhibit soft $E_s(n)$ behaviour), 
the DU threshold is never reached so that no additional constraint for ${\rm v_{ex}}$ arises.
This statement, however, has to be taken with the caveat that we have used a very stiff energy of 
symmetric matter $E_0(n)$ here.
For the $\gamma=9/10$ model the opposite conclusion applies: the DU constraint is violated in all cases except for the very lightest stars and very large excluded volumina.
In order to fulfill the DU constraint for stiff $E_s(n)$ models sufficiently large values of ${\rm v_{ex}}$ have 
to be chosen to keep $n_c$ below $n_{\rm DU}$.
\section{Results and Conclusions}
We have demonstrated on the example of the DD2 EoS that stiffening $E_0(n)$ by increasing the excluded volume correction lowers the central density $n_c$ of a NS with given mass.
We have used the class of MDI-type density dependence of the symmetry energy $E_s(n)$ to show how increasing its stiffness lowers the threshold density $n_{\rm DU}$ for the onset of the DU cooling process.
On this basis, requiring the DU constraint to be fulfilled, i.e. that for typical neutron stars (not exceeding a certain mass between $1.4$ and $1.6~$M$_\odot$) the DU cooling process will not be operative, 
a condition for the minimal stiffness of the of the symmetric EoS for a given symmetry energy 
(above the mass dependent threshold value) has been obtained, shown in Fig.~\ref{vx-nc}.  
For the example  considered in this contribution, there is an approximately linear dependence of the minimal NS radius $R$ (or excluded volume $\mathrm{v_{ex}}$) and the symmetry energy $S$ (or the $\gamma$ parameter). 

Below the NS mass dependent threshold value for the symmetry energy $S$ the DU process will not operate even for the softest EoS in the sample and consequently there is no constraint on the excluded volume $\mathrm{v_{ex}}$.
It is important to note that for softer MDI-type $E_s(n)$ for which $\gamma<0.74$, the fast DU cooling is never activated in typical stars with masses not exceeding $1.6~$M$_\odot$. 
This is due to the fact that the symmetric EoS $E_0(n)$ for  DD2 is already sufficiently stiff in order to not allow densities above $n_{\rm DU}$ in the center of those stars.
\begin{figure}
\begin{center}
\includegraphics[width=0.8\textwidth, angle=0]{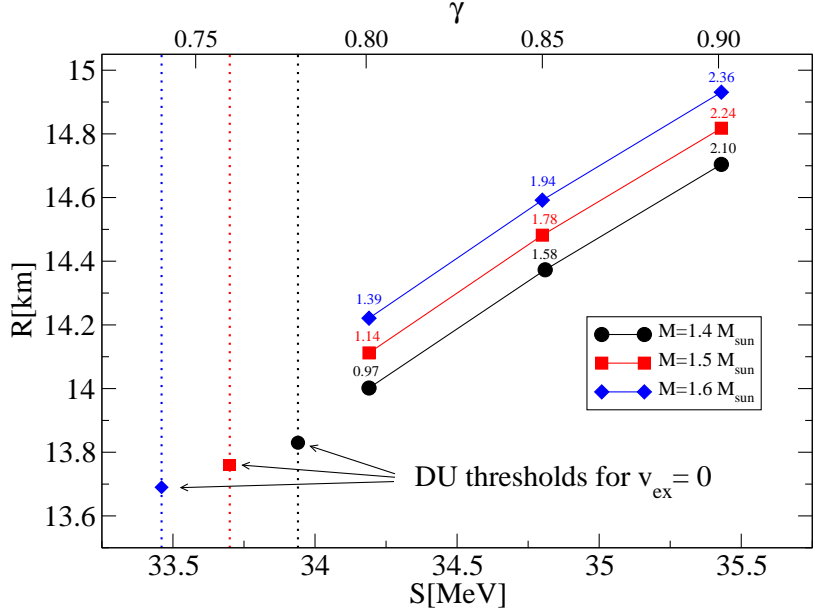}
 \caption{
Effect of the direct Urca cooling constraint on neutron stars of different masses: $1.4~$M$_\odot$ (filled circles), $1.5~$M$_\odot$ (filled squares), and $1.6~$M$_\odot$ (filled diamonds) for the MDI-type parametrization of the symmetry energy and the DD2 parametrization of the symmetric EoS.
Below the values for the symmetry energy $S$ (or $\gamma$) indicated by vertical dotted lines and denoted as "DU thresholds for ${\rm v_{ex}}=0$" no constraints for radii $R$ (or ${\rm v_{ex}}$) arise.
The filled symbols connected with solid lines to guide the eye denote minimal radii $R$  to fulfill the DU cooling constraint; numbers at the symbols denote values of ${\rm v_{ex}}$ in fm$^3$.
\label{S-R}}
\end{center}
\end{figure}

The analysis of excluded volume effects carried out in this contribution complements recent  
studies 
where it was shown that:

\begin{itemize}
 \item a stiffening of the EoS (which leads to a flattening of the 
density profile, as we show in Figure~\ref{M-R-nc}, left panel) 
would lead to a slowing down of the other (non-DU) cooling processes, 
unless this effect is 
compensated by modifications of the density dependence of pairing gaps~\cite{Grigorian:2016leu}. 
\item a stiffening of the EoS (by excluded volume) allows an early onset of deconfinement, 
with central densities as low as $2.4~n_0$.
(albeit at $M_{\rm onset}\sim 2~$M$_{\odot}$) with a possibility for a disconnected stable hybrid star branch (third family, including high- mass twins) \cite{Benic:2014jia,Alvarez-Castillo:2016oln}. 
\item the quark matter phase transition may thus be another solution
of the DU problem whenever it occurs~\cite{Klahn:2006iw}, 
similar to the analoguous solution of the hyperon puzzle by~\cite{Baldo:2003vx}.
\end{itemize}

We would like to remark that the above conclusions are made for the MDI-type symmetry energy functional and therefore model dependent.
Regarding the $E_s(n)$ functionals, we would like to refer to other previous works that derive predictions of NS radii once $E_s(n)$  is measured accurately for a broader density range around saturation~\cite{Steiner:2015aea,AlvarezCastillo:2012rf}.

There is the alternative possibility that the symmetry energy of nuclear matter has a high-density behaviour that a priori does not allow the DU process to occur.
This is the case, for instance, for the class of DD2-type symmetry energies introduced in \cite{Typel:2014tqa}. 
For such $E_s(n)$ it has been demonstrated that the symmetry energy contribution to the NS EoS behaves {\it universal} \cite{Klahn:2006ir}.  
For a recent detailed discussion of this conjecture, see \cite{Blaschke:2016}.
This would allow to extract the symmetric EoS $E_0(n)$ from a sufficiently precise measurement of the neutron star EoS that could eventually be provided by future satellite missions such as NICER.

\section*{Acknowledgments}
We thank the organizers of \textit{The Modern Physics of Compact Stars 2015} for their support and hospitality. We are grateful to Stefan Typel for discussions and collaboration on the excluded volume topic and for providing the data of the excluded volume modified DD2 EoS. 
The work of D.B. has been supported by NCN under grant No. 2014/13/B/ST9/02621. 
D.E.A-C. acknowledges support from the COST Action MP1304 "NewCompStar" as well as from the Bogoliubov-Infeld (for scientific exchange between JINR Dubna and Polish Institutes) and 
Heisenberg-Landau (for exchange between JINR Dubna and German Institutes) programmes.

\end{document}